# Highly Sensitive and Compact Quad-Band Ambient RF Energy Harvester

R. Keshavarz, and N. Shariati, *IEEE Member*

*Abstract*— A highly efficient and compact quad-band energy harvester (QBEH) circuit based on the extended composite right- and left-handed transmission lines (E-CRLH TLs) technique is presented. The design procedure based on E-CRLH TLs at four desired frequency bands is introduced to realize a quad-band matching network (QBMN). The proposed QBEH operates at four frequency bands: $f_1 = 0.75 \text{ GHz}, f_2 = 1.8 \text{ GHz}, f_3 = 2.4 \text{ GHz}$ and $f_4 = 5.8 \text{ GHz}$. The simulations and experimental results of the proposed QBEH exhibit overall (end to end) efficiency of 55% and 70% while excited at four frequency bands simultaneously with $-20 \text{ dBm}$ (10 μW) and $-10 \text{ dBm}$ (100 μW) input power, respectively. Due to applying multi-band excitation technique and radio frequency (RF) combining method in the QBEH circuit, the sensitivity is improved, and sufficient power is generated to realize a self-sustainable sensor ($S^3$) using ambient low-level RF signals. A favorable impedance matching over a broad low input power range of $-50$ to $-10$ dBm (0.01 to 100 μW) is achieved, enabling the proposed QBEH to harvest ambient RF energy in urban environments. Moreover, an accurate theoretical analyses based on the Volterra series and Laplace transformation are presented to maximize the output DC current of the rectifier over a wide input power range. Theoretical, simulation and measurement results are in excellent agreement, which validate the design accuracy for the proposed quad-band structure. The proposed new energy harvesting technique has the potential to practically realize a green energy harvesting solution to generate a viable energy source for low-powered sensors and IoT devices, anytime, anywhere.

*Index Terms*— Ambient Energy harvesting, electromagnetic energy, high efficiency, IoT, quad-band rectifier, rectenna (rectifying antenna), wide input power range.

## I. INTRODUCTION

The Internet of Things (IoT) is spawning the use of sensors, which are both pervasive and ubiquitous in every aspect of our lives. Whilst recent advances in modern sensors and IoT devices appear to be unlimited, the dependence of their operation on batteries remains a significant weakness which has imposed restrictions on the range, lifetime, environmental pollution, and are often challenging to replace them [1]-[3]. This creates an inspiration to develop an alternative method to feed sensors. Two promising approaches are wireless power transfer (WPT) [4]-[6] and energy harvesting from ambient sources [7]-[11].

Energy harvesting refers to collecting energy from wind, sun, thermal, water, and RF signals generated by wireless communications systems [12]-[15]. Electromagnetic energy harvesting is a green and cost-effective energy solution to provide a sustainable energy source for portable IoT devices, remotely deployed self-sustainable sensors ($S^3$), Wireless Sensor Networks (WSNs), and wearable devices. Moreover, widespread applications of self-sustainable sensors in many areas including agriculture, mining, health monitoring, and smart cities, has triggered impetus for energy harvesting [6], [16].

Real-world measurements have shown that typical propagated power levels from a GSM base station or a WLAN access point reach several $\mu W/cm^2$ within 20–200 m, especially in urban areas. These low-energy signals can provide a green power source to enhance the life time of ubiquitous devices through continuous ambient energy scavenging [17]. Power management and energy storage (supercapacitor or battery) could be integrated into the energy harvesting system to store captured energy, leading to provide a stable output voltage.

The prominent advantage of RF harvesting is the ability to convert electromagnetic energy into electricity during day and night, both indoor and outdoor. Furthermore, penetrations of RF signals inside the structure (e.g., walls, bridges, tunnels), and underground allow for electromagnetic (EM) energy harvesting where other energy sources (e.g., solar, wind) are not available. Further, a combination of EM energy harvesting and other available energy sources (solar, thermal, piezo, etc.) can develop a hybrid energy harvesting solution with efficient DC combining strategy.

The majority of published papers are mainly focused on a specific frequency band, such as GSM [21], Bluetooth [22], Wifi bands [23] and at a given input power level. However, if ambient frequency shifted, the functionality of such harvesters will be significantly affected. Furthermore, energy harvester systems with tunable impedance matching networks are not widely applicable for environmental RF energy scavenging, where available power is very low [24].

The available ambient RF power levels can vary unpredictably, depending on the distance from the transmission source, the transmission media, and the antenna orientation. Therefore, from an ambient energy scavenging perspective, the efficiency, and sensitivity of an energy harvester (EH) are key factors [18]. Enabling simultaneous multi-band and multi-tone signals in the input of an EH and taking the advantage of RF combining method, the rectification device can be turned on at low input power levels available in ambient, hence the rectifier sensitivity can be improved, resulting in enhanced output DC power, provided that the system can operate over a broad input power range and at different frequency bands [19], [20].

Due to the non-linearity of rectification device (e.g., diode), variation of rectifier input impedance over a broad frequency range is very large [24]. Therefore, for simplicity, using separate EH modules consisting of single-band rectennas in parallel branches is proposed, where several frequency bands are available. In this approach, each branch is tuned at a single frequency band and a DC combiner is used to accumulate the

RF and Communication Technologies (RFCT) research laboratory, University of Technology Sydney, Ultimo, NSW 2007, Australia, e-mail: Rasool.Keshavarz@uts.edu.au; Negin.Shariati@uts.edu.au

DC power produced by all single modules [25], [14]. However, using multiple narrowband antennas for each branch leads to a bulky system that is not practical for low profile devices. On the other hand, using a wideband antenna for all branches needs a power splitter and a bank filter to feed each branch at the specified frequency band, so RF loss increases drastically [25]. As a result, the parallel branches are not capable of capturing low-level ambient RF signals, hence this solution is not suitable for ambient RF energy scavenging.

Another approach to design a multi-band EH is using a wideband antenna and a multi-band matching network to enhance RF combining at the input of the rectifier circuit [26]-[28]. In this method, RF waves in multi-band signals excite the rectification device (e.g. diode) simultaneously to improve the EH sensitivity. The major challenge in this method is designing a multi-band matching network to match the antenna output to the rectifier in all operational frequency bands and over a wide input power range [28].

In the past decade, a novel extended composite right/left-handed (E-CRLH) transmission line (TL) metamaterial (MTM) structure is proposed [29] and recently dual CRLH (D-CRLH) prototypes are introduced [30]. The family of multi-band metamaterial-inspired matching networks in EH systems is reported based on several types of metamaterial configurations [31]-[33]. Quad-band RF devices using E-CRLH structure can be realized which are suitable in designing quad-band matching network in EH systems [34]-[36].

Moreover, many attempts were made on nonlinear analysis of an EH system, and its operation under various input waveforms [25] and [37]. Nevertheless, these studies used a simplified nonlinear model of a diode in the mathematical analysis which is not practical. Likewise, a closed-form equation for output voltage (or current) of the load impedance was not provided.

In this paper, a compact quad-band energy harvester (QBEH) is proposed using E-CRLH TL as a quad-band matching circuit. Due to its compactness and low-cost properties, this structure has a potential to be used as an electromagnetic energy harvester module in low-power IoT devices and wireless sensor networks.

Contributions of this paper are summarized as follows:
- For the first time, we designed and analyzed a rectifier topology with four concurrent input signals at four specific frequency bands through a quad-band matching network. The proposed highly sensitive rectifier is matched over a wide low input power range of −50 to −10 dBm (0.01 μW to 100 μW), which is very challenging due to the nonlinearity of the rectification device with input power [18].
- As a result of combining four signals simultaneously into the rectifier, the proposed QBEH exhibits higher sensitivity and overall (end-to-end) efficiency compared to previously published papers.
- Using multi-tone excitation in each frequency band enhances the output DC power level of an EH [20]. Therefore, the effect of simultaneous multi-tone excitation on the proposed QBEH is investigated.
- From a design and economic perspective, utilizing a large number of components (e.g. antennas, diodes) to realize individual rectifier circuits for each frequency band results in additional expense, and increasing the total footprint. Thus, the proposed energy harvester is compact and low cost and has the potential to be integrated into various industrial sensors and IoT systems with an antenna and supercapacitor/battery to realize a rectenna as an alternative energy source.
- The design guide procedure for arbitrary load impedance and operational frequency band has been presented and equations were derived.
- A complete nonlinear equivalent circuit model of the diode is used to develop an accurate theoretical model and a comparison between theory, simulation, and measurement is presented.
- This work for the first time demonstrates a new theoretical nonlinear analysis for the proposed QBEH circuit based on Laplace transformation and Volterra series to derive closed-form equations for output DC current ($I_{DC}$). The proposed analysis aims to choose optimum values for the rectifier components to achieve maximum $I_{DC}$ through a wide input power range at four frequency bands. The theoretical analysis of section III is explained in detail in Appendix.
- Furthermore, applying Volterra series to a set of nonlinear equations [38] and deriving closed-form equation for $I_{DC}$ is very complicated while considering all components of the rectifier circuit together (including two diodes, two capacitors, and load). In this paper, the classical series is applied to the Laplace domain of the nonlinear equations instead of applying to the nonlinear equations themselves and hence deriving closed-form is relaxed.

The organization of this paper is as follows: Step by step design procedure and equations of the quad-band rectifier are presented in section II. Section III and Appendix demonstrate theoretical analyses of the proposed quad-band energy harvester (QBEH). The QBEH performance is validated by analytical, simulation, and measurement results in section IV. Finally, conclusions are provided in section V.

## II. QUAD-BAND RECTIFIER DESIGN METHODOLOGY

Ambient EH circuit scavenges RF signals from freely available sources in the environment and converts them into useful DC power using a rectifying antenna (rectenna). A typical rectenna consists of five main sections: receiving antenna, matching network, voltage-doubler rectifier circuit, load impedance ($Z_L$) and storage (super-capacitor/battery) [39] (Fig. 1). Furthermore, load impedance may include a capacitor to remove output ripples of the rectifier as a low pass filter.

According to Fig. 1, when RF signal is received by antenna and after filtering, the input impedance of the rectifier is matched to the reference impedance of the system (50 Ω) using a matching network. Based on the maximum power transfer theorem, when the source and load impedances are complex conjugates, the highest amount of energy transfers to the load. In the design of an RF to DC power conversion circuit, a voltage-doubler structure is used as this topology is well suited to low power rectification [39]. The output DC power collected into the load can be stored in a supercapacitor or battery.

Many works presented single-band electromagnetic energy harvesting circuits [40], [22]. However, if the operating frequency shifted, the functionality of such harvesters is severely degraded. Tuning the load impedance ($Z_L$) is also proposed to enhance the efficiency [24]. However, by tuning the load, the agility of the harvester is decreased. Further, the main goal in designing an efficient RF harvesting system is to produce high output DC power for fixed loads such as sensors. Therefore, a multi-band EH benefits from the accumulation of RF signals in several frequencies, and hence, a higher amount of energy can be harvested [26], [31].

In this paper, 'multi-band' denotes continuous wave (CW) signals with different frequency bands, and 'multi-tone' refers to multi-sine excitation in each frequency band [41]. In the proposed multi-band EH circuit, the BPF and matching network merged to realize a quad-band matching network (QBMN) (Fig. 2).

A quad-band device is capable of accomplishing similar operation at four different arbitrary frequencies $f_1, f_2, f_3$ and $f_4$. Such a device is therefore constituted of TL sections inducing equivalent phase shifts ($\theta_i$):

$$\theta_i = \beta_i p \qquad i = 1,2,3 \text{ and } 4 \tag{1}$$

where $p$ is the physical length of a unit cell and $\beta_i$ is the propagation constant in each frequency band.

The unit cell is realized by combining a conventional CRLH, its dual structure (D-CRLH), and an extended CRLH (E-CRLH). Figure 3 shows the proposed unit cell of an E-CRLH TL which is used as QBMN (quad-band matching network) in Fig. 2. This circuit consists of eight lumped components which lead to four $LC$ resonances; two series $LC$ resonances ($\omega_{cs}$, $\omega_{dp}$) and two parallel $LC$ resonances ($\omega_{cp}$, $\omega_{ds}$). It is also known that E-CRLH TL can operate as a quad-band structure [35]. In this work, eight equations were derived to synthesize a quad-band transmission line with E-CRLH cells. From these equations, the values of lumped elements in Fig. 3 can be determined.

Considering Fig. 3, series impedance and parallel admittance of an E-CRLH unit cell are derived as follows:

$$Z = j\omega L_R^c \left(1 - \left(\frac{\omega_{cs}}{\omega}\right)^2\right) - \frac{j}{\omega C_L^d \left(1 - \left(\frac{\omega_{dp}}{\omega}\right)^2\right)} \tag{2}$$

$$Y = j\omega C_R^c \left(1 - \left(\frac{\omega_{cp}}{\omega}\right)^2\right) - \frac{j}{\omega L_L^d \left(1 - \left(\frac{\omega_{ds}}{\omega}\right)^2\right)} \tag{3}$$

Where

$$\omega_{cs} = \frac{1}{\sqrt{L_R^c C_L^c}}, \ \omega_{cp} = \frac{1}{\sqrt{L_L^c C_R^c}}, \ \omega_{dp} = \frac{1}{\sqrt{L_R^d C_L^d}}, \ \omega_{ds} = \frac{1}{\sqrt{L_L^d C_R^d}} \tag{4}$$

The propagation constant and the characteristics impedance in terms of the transmission matrix elements of the E-CRLH TL unit cell are equal to [35]:

$$\cos(\beta p) = 1 + ZY/2 \tag{5}$$

$$Z_c \cong \sqrt{\frac{Z}{Y}} \tag{6}$$

Diode is key part of the loading impedance in QBMN design. A low forward-voltage drop and fast switching Schottky diode is selected in the design of the proposed QBEH circuit. The sensitivity of an energy harvester improves when its forward-voltage drop is very low. Consequently, in this work, the Schottky diode HSMS2850 is used in the rectifier circuit, and spice parameters are presented in Table I [42].

According to Fig. 2, load impedance of the output port consists of two diodes ($D_1$, $D_2$), two capacitors ($C_1$, $C_2$), and load impedance ($Z_L$). The loading effect of these components are considered in the QBMN design procedure, which is shown with $Z_L$.

According to Fig. 2 and Fig. 3, the QBMN input impedance is:

$$Z_{in} = \frac{1}{\frac{1}{\frac{Z}{2}+Z_L}+Y} + \frac{Z}{2} \tag{7}$$

From (5), (6), and (7), for the first time, the following polynomial equation of degree four is readily established.

$$\left(\frac{(Z_L-Z_{in})\cos(\theta)}{Z_L Z_{in}-\frac{Z_c^2}{2}(1+\cos(\theta))}\right)^2 + \frac{2(1-\cos(\theta))}{Z_c^2} = 0 \tag{8}$$

A proportional phase difference for a determined $Z_{in}$ and $Z_L$ values in all four frequency bands are calculated and the lumped components of ECRLH TL are defined [35]. In the design procedure, $Z_L$ can be derived from simulation results for $Z_{in} = 50\ \Omega$. Finally, $\theta$ in each frequency band is determined using equation (8) in MATLAB software. Moreover, a design guide is presented for four arbitrary operational frequency bands to achieve a quad-band matching network based on required conditions.

### III. THEORETICAL ANALYSIS OF QUAD-BAND RECTIFIER

Step by step design procedure of the proposed quad-band matching network (QBMN) is discussed. This section aims to provide the most accurate theoretical analysis of the proposed QBEH to obtain optimum components values in the rectifier circuit and to match over a wide input power range at four frequency bands, leading to maximum DC current at the load. In this section, firstly, nonlinear equations of the complete rectifier circuit with two diodes are mentioned. Then, as solving a set of nonlinear equations and deriving closed-form for DC load current is complicated, the Volterra series technique will be applied to the Laplace transformation of the nonlinear equations, not to the nonlinear equations themselves. This approach simplifies the set of equations and a closed-form of load current will be derived. Details of the proposed analysis are given below.

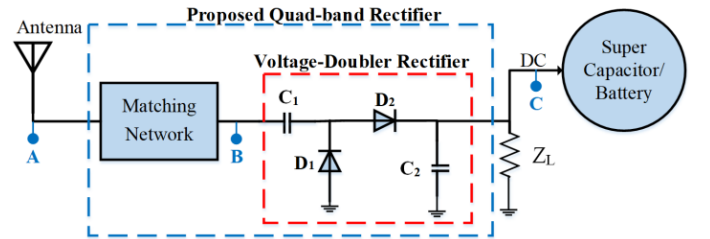

Fig. 1. Typical ambient EH system.

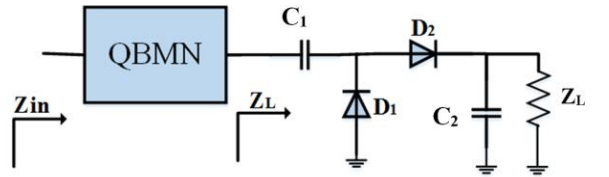

Fig. 2. Schematic of the proposed quad-band EH circuit.

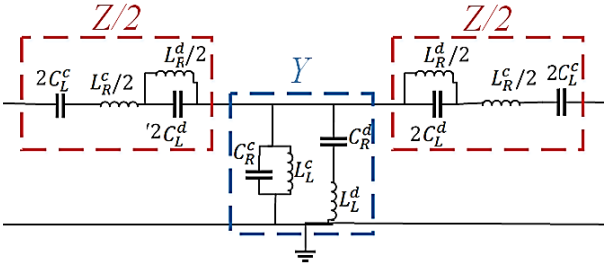

Fig. 3. Schematic of the E-CRLH unit cell as a quad-band matching network.

TABLE I
SPICE PARAMETERS OF HSMS2850

| $B_V(V)$ | $C_{j0}(pF)$ | $E_G(eV)$ | $I_{BV}(A)$ | $I_S(A)$ | $N$ | $V_J(V)$ | $R_s(\Omega)$ |
|---|---|---|---|---|---|---|---|
| 3.8 | 0.18 | 0.69 | 3×10E-4 | 3×10E-6 | 1.06 | 0.35 | 25 |

The equivalent circuit model of the proposed rectifier is illustrated in Fig. 4. A nonlinear model of the Schottky diode is considered, including a nonlinear conductance $G(v_j)$ in parallel with a nonlinear capacitance $C(v_j)$ and a fixed resistance $R_s$ (Fig. 4). The Schottky diode is represented by nonlinear equations [43]:

$$i_{D1} = I_s \exp(\alpha v_{j1}) - \frac{C_0}{\sqrt{1-\frac{v_{j1}}{V_0}}} \frac{dv_{j1}}{dt} \quad (9)$$

$$i_{D2} = I_s \exp(\alpha v_{j2}) - \frac{C_0}{\sqrt{1-\frac{v_{j2}}{V_0}}} \frac{dv_{j2}}{dt} \quad (10)$$

where $C_0$ is the static capacitance of the Schottky set by DC bias $V_{DC}$, $I_s$ is the saturation current and $\alpha = \frac{1}{nV_T}$ ($V_T = \frac{kT}{q}$ is the thermal voltage, $n$ is the diode ideality factor, $k$ is the Boltzman constant, $T$ is the junction temperature, and $q$ is the electron charge).

Based on the Taylor expansion, equations (9) and (10) can be represented as:

$$i_{D1} = I_s\left(1 + \alpha v_{j1} + \frac{\alpha^2}{2}v_{j1}^2 + \frac{\alpha^3}{6}v_{j1}^3 + \cdots\right) + C_0\frac{dv_{j1}}{dt}\left(1 + \frac{1}{2V_0}v_{j1} + \frac{3}{8V_0^2}v_{j1}^2 + \frac{5}{16V_0^3}v_{j1}^3 + \cdots\right) \quad (11)$$

$$i_{D2} = I_s\left(1 + \alpha v_{j2} + \frac{\alpha^2}{2}v_{j2}^2 + \frac{\alpha^3}{6}v_{j2}^3 + \cdots\right) + C_0\frac{dv_{j2}}{dt}\left(1 + \frac{1}{2V_0}v_{j2} + \frac{3}{8V_0^2}v_{j2}^2 + \frac{5}{16V_0^3}v_{j2}^3 + \cdots\right) \quad (12)$$

Now, applying Kirchhoff's laws to the proposed circuit in Fig. 4, equation (13) can be derived which consists of three sub-equations:

$$\begin{cases}(C_1R_sD + 1)i_{D1} - i_{D2} - C_1D\ v_{j1} + C_1D\ v_i = 0 \\ i_{D2} - (C_2\ R_L D + 1)i_L = 0 \\ R_s i_{D1} + R_s i_{D2} + v_{j1} + v_{j2} + R_L i_L = 0\end{cases} \quad (13)$$

Where

$$D = \frac{d}{dt} \quad (14)$$

and unknown variables are:

$$[i_{D1}\ i_{D2}\ v_{j1}\ v_{j2}\ i_L]^T \quad (15)$$

Therefore, a set of five equations are achieved; three linear (13), and two nonlinear equations (11) and (12), which cannot be solved easily. Hence, to continue the nonlinear analysis of the proposed QBEH circuit and to derive a closed-form equation for output DC current ($I_{DC}$), Laplace transformation and Volterra series are used. $I_{DC}$ can be readily determined as shown in Appendix:

$$I_L(DC) = \lim_{s\to 0}\left[\frac{C_1s^2F_2(s).V_i(s)+C_1s^2F_3(s).B(s)+C_1s^2A(s)}{C_1sF_1(s)}\right] \quad (16)$$

and

$$V_L(DC) = R_L.\ I_L(DC) \quad (17)$$

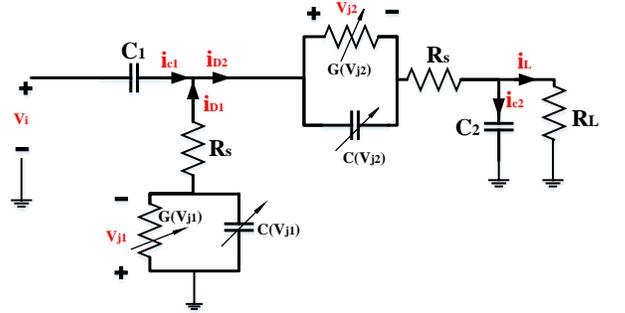

Fig. 4. Equivalent circuit model of the proposed voltage-doubler rectifier.

Finally, according to (16) and (17) and using MATLAB, the optimum values of the rectifier components will be obtained to maximize $I_L(DC)$ over a broad input power range at four desired frequency bands (section IV). Furthermore, these equations can be used as a foundation for more optimization procedures in future works.

## IV. SIMULATION AND EXPERIMENTAL RESULTS

According to Fig. 2, the rectifier input impedance, $Z_L$, is initially defined to design the matching network, while the antenna input impedance is selected as 50 Ω. The nonlinear behavior of the proposed rectifier is modeled using Advanced Design System (ADS) software. Figure 5 illustrates a large-signal analysis of the proposed rectifier which is optimized based on the theoretical analysis of section III. In this figure $Z_L$ varies with frequency and input power (−30<Pin<0 dBm). The quad-band matching network (QBMN) topology is a quad-band E-CRLH bandpass filter (Fig. 3) to match the antenna impedance and $Z_L$ at the selected frequency bands and over a broad low input power range.

Based on the Australian Radiofrequency Spectrum Plan and our previous RF field investigations [44], [14], [10] the frequencies of TV signal (750 MHz), GSM1800 (1800 MHz), Bluetooth/WiFi (2.4 GHz), and WiFi (5.8 GHz) are selected for the proposed quad-band energy harvester (QBEH). These frequencies are allocated to broadcasting services, communications, and cellular systems in Australia [14].

According to the design procedure in section II and ADS simulations, the component values of the quad-band E-CRLH band-pass filter as QBMN are calculated and optimized (Table II). The closest values to these components that are available in practice have been used in fabrication.

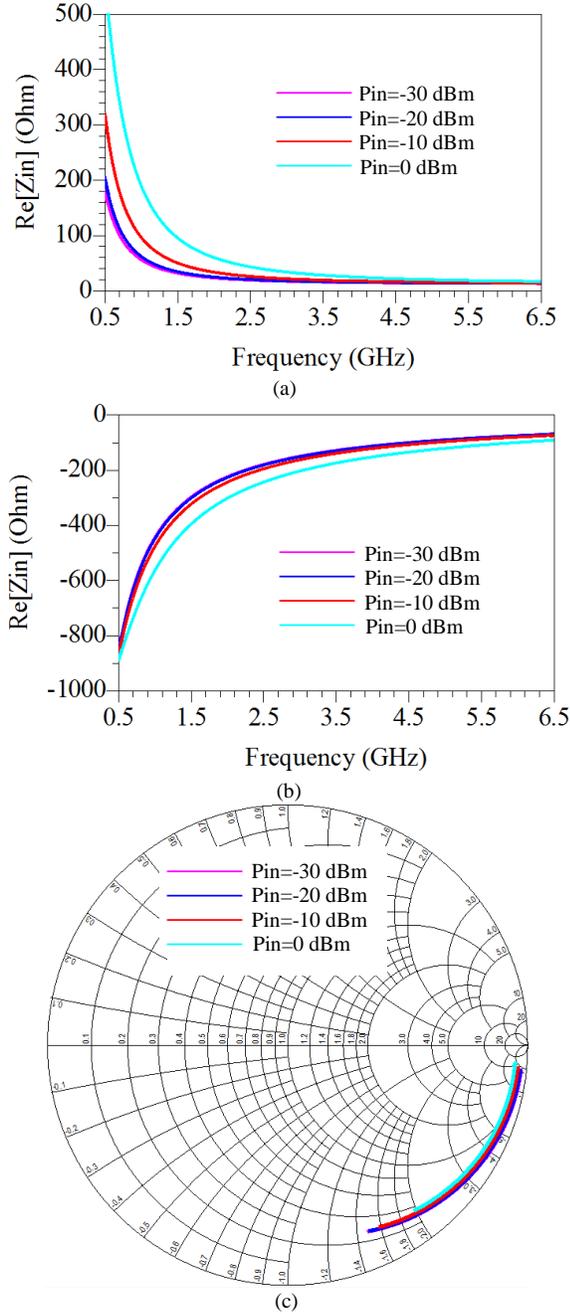

Fig. 5. Input impedance of the proposed rectifier circuit at different input power levels ($-30<\text{Pin}<0$ dBm, step=10 dB) calculated with Large Signal S-parameter analysis over the frequency range of 0.5 to 6.5 GHz, a) real part, b) imaginary part, c) Smith chart.

To evaluate the proposed structure performance, the QBEH circuit is developed based on the derived equations and design procedure (Sections II and III), and simulated. The structure is fabricated on a low-cost FR-4 substrate with $\varepsilon_r = 4.7$, thickness of 1.6 mm, and tan δ = 0.02 (Fig. 6). The QBEH with dimensions of 12×47 mm operates at four frequency bands of $f_1 = 0.75\ GHz, f_2 = 1.8\ GHz, f_3 = 2.4\ GHz$ and $f_4 = 5.8\ GHz$ with fractional bandwidth of 12%, 7%, 3% and 8% respectively.

### A. Reflection Coefficient

The $|S_{11}|$ of the QBMN was evaluated using a PNA Microwave Network Analyzer (N5245B) which was re-calibrated for each input power level.

In Fig.7, the $|S_{11}|$ of the QBMN is presented at low input powers ($P_{in} = -30, -20, -15, and -10\ dBm$) which shows nearly 90 MHz, 120 MHz, 60 MHz and 450 MHz bandwidth around the design frequency bands of $f_1 = 0.75\ GHz, f_2 = 1.8\ GHz, f_3 = 2.4\ GHz$ and $f_4 = 5.8\ GHz$, respectively. The measured results show very good agreement with simulations (the slight difference is due to the diode characteristics). Design goal of the proposed QBMN is to improve the rectifier sensitivity at low input power levels which are available in ambient. Fig. 8 demonstrates the measured S-parameter of the QBEH in both frequency and input power dimensions. The $|S_{11}|$ is a quasi-concave curve and reaches its minimum around four operational frequencies over a broad input power range of $P_{in} = -50\ to\ 0\ dBm$. This curve shows the best matching condition is fulfilled around the design goal of $P_{in} = -30\ dBm$. Due to the non-linearity of the diode impedance with input power and frequency, the input matching is changed over a broad range of input power and frequency. Simulation and measurement results of the reflection coefficient are in great agreement (Fig. 7 and Fig. 8) which validate the accurate design procedure and theoretical analyses of the proposed QBMN (Sections II and III).

In order to evaluate the practicality of the proposed QBEH in ambient EM energy scavenging applications, and also to provide a realistic scenario, a specific criterion is defined. Table III summarizes characteristics of typical sensors that have been used in agriculture and industry. According to this table, a mean value of 11 kΩ is used as a practical load value in simulations and measurements.

TABLE II
OPTIMIZED PARAMETERS OF THE CHIP COMPONENTS FOR THE PROPOSED E-CRLH.

| $C_R^c$ (pF) | $L_L^c$ (nH) | $L_L^d$ (nH) | $C_R^d$ (pF) | $L_R^c$ (nH) | $C_L^c$ (pF) | $C_L^d$ (pF) | $L_R^d$ (nH) |
|---|---|---|---|---|---|---|---|
| 0.04 | 49.4 | 120.7 | 0.04 | 9.6 | 0.2 | 0.48 | 11 |

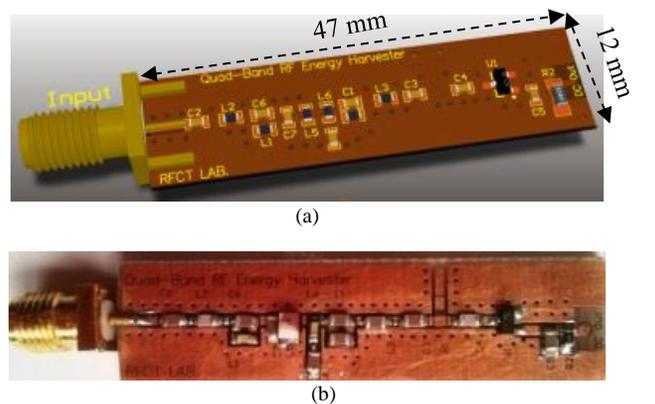

Fig. 6. a) 3D layout, b) Fabricated prototype of the proposed QBEH.

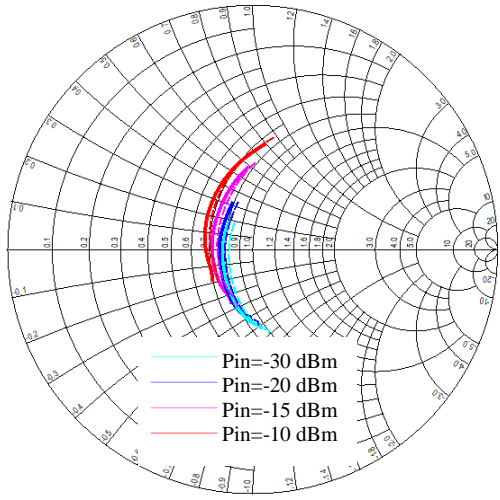

(a)

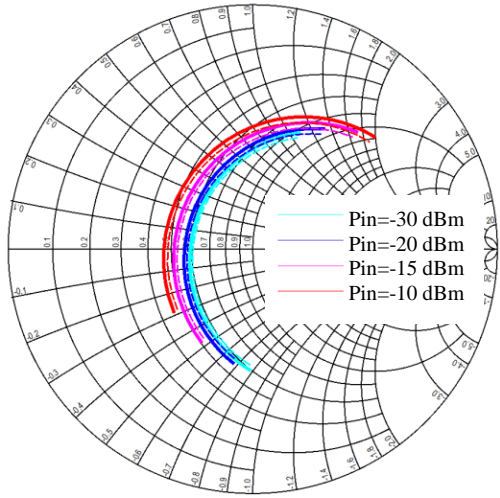

(b)

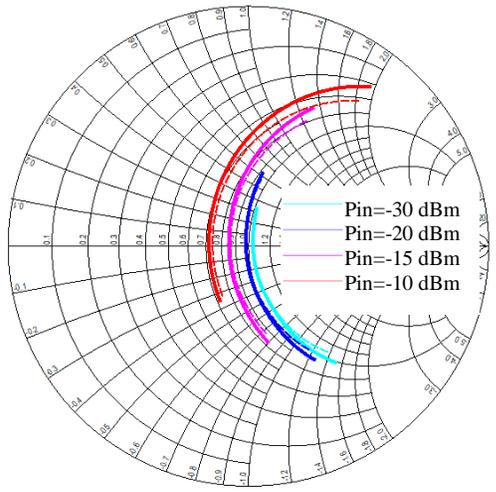

(c)

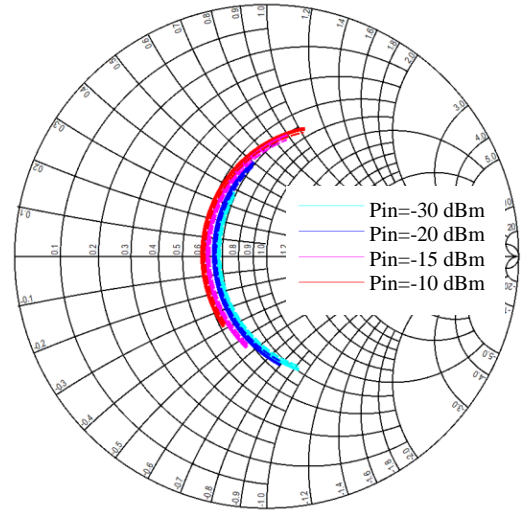

(d)

Fig. 7. Measured (solid line) and simulated (dash line) $|S_{11}|$ of the proposed QBEH at $P_{in} = -30, -20, -15$ and $-10\ dBm$, a) 700~800 MHz, b) 1.7~1.9 GHz, c) 2.3~2.5 GHz, d) 5.3~6.5 GHz

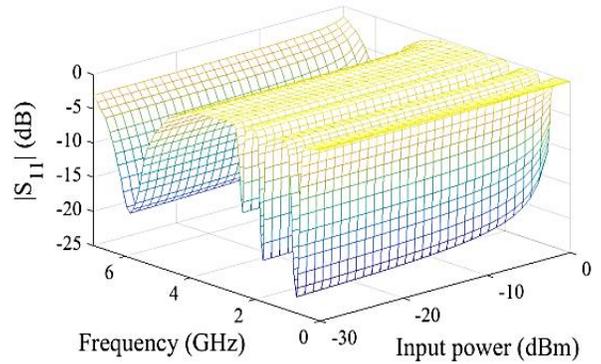

Fig. 8. Measured $|S_{11}|$ of the QBEH as a function of frequency and input power.

TABLE III
CHARACTERISTICS OF TYPICAL SENSORS

| Parameters | [45] | [46] | [47] | [48] | [49] |
|---|---|---|---|---|---|
| DC supply voltage (V) | 1 | 0.3 | 0.8 | 0.45 | 0.3 |
| Power consumption ($\mu W$) | 150 | 10 | 15 | 14 | 3.7 |
| Input impedance (k$\Omega$) | 7 | 9 | 11 | 14 | 24.3 |

### B. Output Voltage and Power

Theoretical, simulation and measurement results of the output DC voltage of the proposed QBEH are presented in Fig.9. Please note, multi-band denotes continuous wave (CW) signals with different frequency bands while multi-tone refers to multi-sine excitations in each frequency band. Measurements were performed using PSG Analog Signal Generators (E8257D) as RF power sources and recording of the output DC voltage across the load resistance was achieved with a Fluke digital voltage meter.

Fig. 9.a shows measured output DC voltage of the QBEH in single-band and quad-band scenarios. This figure exhibits that the QBEH operates properly in each frequency band and obviously, DC output voltage of the QBEH in quad-band case is more than single-band. Moreover, roughly similar output DC voltages in each four frequency bands are achieved, which validate the steady performance of the proposed QBMN based on E-CRLH TL. Furthermore, due to the non-linearity of diode

with frequency and input power, there is always a slight difference between output voltages, despite of matching condition (Fig. 9.a). According to the datasheet of HSMS2850, the diode output voltage at 915 MHz, 2.45 GHz and 5.8 GHz are nearly similar, if the matching condition is fulfilled in these frequency bands [42].

Calculated, simulated, and measured output DC voltage of the proposed QBEH in single-band and quad-band scenarios are presented in Fig. 9.b, which is focused on the lower power section of the input power ranging from 0 to $20\ \mu W\ (-17\ dBm)$. In the single-band input signal, the curve shows an average output voltage of four frequency bands. As can be seen, 200 mV output voltage is produced with 12 µW input signal in the single-band mode, while only 2 µW input signal is required to produce the same amount of output DC voltage in the quad-band case. These results confirm the accuracy of the proposed nonlinear analysis method and simulation results. The measured output DC power of the proposed QBEH circuit is presented in Fig.10.a, where single-band and quad-band scenarios are compared.

Fig. 10.b compares theoretical, simulation and measurement output power results in the lower input power range of 0 to $20\ \mu W\ (-17\ dBm)$. In this figure, the single-band curve displays an average output DC power in four frequency bands. The measurement results again validate the accuracy of the proposed nonlinear analysis method and simulation results.

According to Fig. 10.a, the minimum required input power to activate the quad-band EH is lower than the single-band, which validates the sensitivity improvement by using a QBEH structure. Measurement results demonstrate that applying a quad-band input to the QBEH circuit can generate nearly 6 times more output DC power compared to the single-band case at $P_{in} = -20\ dBm$ (10 µW). It should be highlighted that producing over 4 times fold increase in the output power of the quad-band structure is due to the non-linearity behavior of diodes which is investigated in section III. It is evident that the proposed quad-band technique increases the RF to DC conversion efficiency, and hence the recoverable DC power for low power applications. Furthermore, the output power enhancement due to RF combining is more effective at the lower power levels, which is the main goal in designing a highly sensitive and efficient electromagnetic energy scavenger.

The achieved sensitivity of the proposed QBEH is improved significantly especially at low input power levels compared to single-band energy harvesters, which demonstrates the usefulness of RF combining method in the proposed circuit.

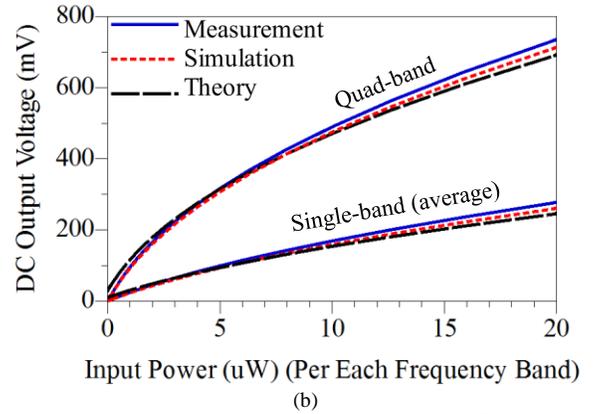

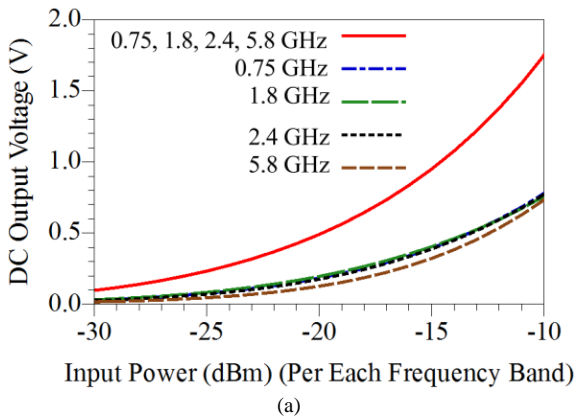

Fig. 9. a) Measured output DC voltage as a function of input RF power for single-band input signal at 0.75 GHz, 1.8 GHz, 2.4 GHz, and 5.8 GHz as well as for the quad-band input signal at input power levels ranging from $-30\ to\ -10\ dBm$ (1 µW to 100 µW), b) Theoretical, simulation, and measurement results of the proposed QBEH over low power region of 0 to $20\ \mu W\ (-17\ dBm)$.

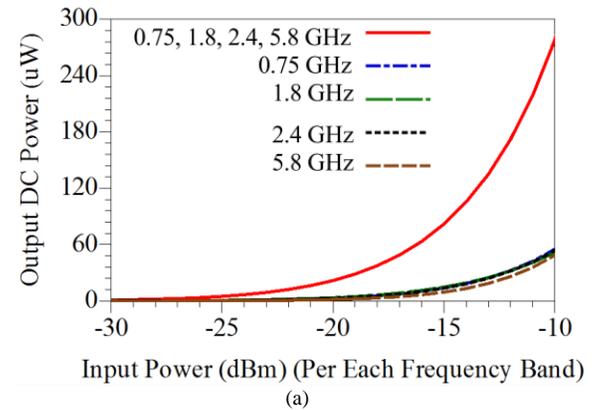

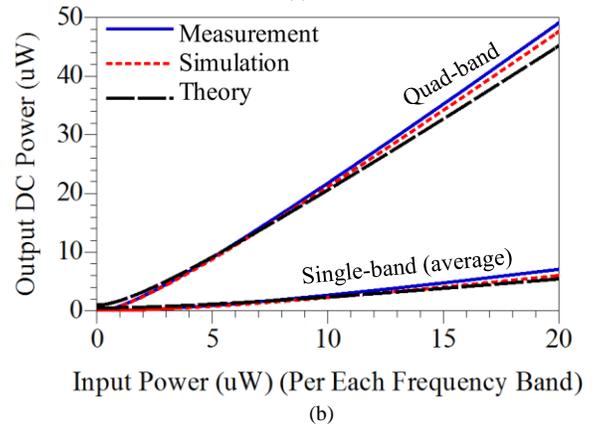

Fig. 10. a) Measured output DC power as a function of input power for four-tone and single-tone excitations at the input power range of $-30\ to\ -10\ dBm$ (1 µW to 100 µW), b) Theoretical, simulation, and measurement results comparison of the proposed QBEH in lower power region of 0 to 20 µW ($-17\ dBm$).

For example, with single-tone excitation at each frequency band, 20 µW input power is sufficient to supply DC voltage of typical sensors (refer to [46], [47], [48], [49] in Table III). As can be seen in Fig. 10.b, this power can be provided using the proposed QBEH when input power at each frequency band is only $10\ \mu W\ (-20\ dBm)$. Whereas the sensors' required power cannot be generated with the low input power level of 10 µW ($-20\ dBm$) in the single-band mode. The sensors would be turned-on in the single-band structure with the minimum input power of 50 µW ($-13\ dBm$).

The required input power can be further reduced in the multi-band structure, by applying multi-tone excitations in each frequency band which will be discussed in section *D*.

### C. Efficiency

Efficiency is one of the key characteristics and metrics of an energy harvesting system. Several definitions of efficiency for multi-band structures have been proposed [18].
According to Fig. 1, the EH system consists of an antenna with the efficiency of $\eta_{Ant}$ and an EH circuit. The circuit section of the EH system consists of matching network, rectifier and load. The efficiency for each part of the circuit can be defined as:

Matching network efficiency: $\eta_{MN} = \frac{P_B}{P_A}$ (18)

Rectifier efficiency: $\eta_{Rec} = \frac{P_C}{P_B}$ (19)

Hence, two overall (end-to-end) efficiencies can be expressed: overall efficiency of the circuit; point *A* to *C* ($\eta_{oc}$) and overall efficiency of the EH system or rectenna efficiency, ($\eta_{oS}$) as:

$\eta_{oc} = \eta_{MN} \times \eta_{Rec} = \frac{P_{dc}}{\sum_{i=1}^{n} P_{RF,i}}$ (20)

$\eta_{oS} = \eta_{oc} \times \eta_{Ant}$ (21)

Where $P_{dc}$ is the total output DC power, $P_{RF,i}$ is the input RF power in the *i*th frequency band right after the antenna output (point *A* in Fig. 1) and *n* is the number of frequency bands.
This paper focuses on the circuit design, analysis, and implementation of a quad-band rectifier and hence, $\eta_{oc}$ is the key metric for comparison with other papers. In some references which combined individual single-band modules in parallel branches to realize a multi-band EH using a bank filter after the antenna, the efficiency of the bank filter, $\eta_{BPF}$ must be considered [25]. Further, in some papers on multi-band EH circuits, the effective efficiency ($\eta_{effective}$) is defined as the ratio of output DC power to the available input RF power in each frequency band, rather than delivered power to the circuit [39]:

$\eta_{effective} = \frac{Total\ output\ DC\ power}{Input\ RF\ power\ in\ ith\ band} = \frac{P_{dc}}{P_{RF,i}}$ (22)

According to (22), the effective efficiency of the proposed QBEH system for a single-tone input power of −30 dBm (1 μW) in each band is equal to 80%, while in [39] the effective efficiency is nearly zero at the same input power for a dual-band structure. In this paper, two above efficiencies ($\eta_{effective}, \eta_{oc}$) are defined to highlight the performance of this work compared to other references. Further, $\eta_{oc}$ is presented in Table IV.

According to equation (20), Fig. 11.a presents the overall efficiency of the QBEH circuit in quad-band and single-band modes. The efficiency is over 55% at $P_{in} = -20\ dBm\ (10\ \mu W)$ in each frequency band where total input power of $-14\ dBm$ is delivered to the input port of the QBEH circuit (point A in figure 1). Also, 70% efficiency is achieved at $P_{in} = -10\ dBm\ (100\ \mu W)$ for the total combined input power of $-4\ dBm$ at the input of the QBEH. Fig. 11.b shows theoretical, simulation, and measurement results of the overall efficiency (equation (20)) for quad-band and single-band modes.

For example, in a quad-band case, the overall efficiency of the proposed QBEH at $20\mu W\ (-17\ dBm)$ input power in each frequency band is 60% which is nearly 1.7 times of the single-band efficiency. It is clear that the proposed quad-band technique enhances the RF to DC conversion efficiency and recoverable DC power. The overall efficiency improvement, especially in the low input power range, is due to implementing RF combining technique in the quad-band structure.

### D. Multi-tone Excitations

As the number of RF sources increases, the generated stacked output DC power increases which will be eventually saturated. A set of concurrent multi-tone signals in each frequency band is applied to the QBEH and simulation results are presented in Fig. 12. As can be seen, the output DC power of the proposed quad-band rectifier with four-tone excitation at each band (total 16 tones), is more than 4 times of single-tone excitation in each band (4 tones totally) at $Pin=20\ \mu W\ (-17\ dBm)$. Consequently, generated output DC power of the sensitive QBEH is saturated by applying multi-tone excitation with over 12 concurrent tones in each frequency band (48 tones totally) at the low input power of $Pin=5\ \mu W$.

In order to further investigate the sensitivity improvement of the proposed QBEH, multi-tone excitations analyses are performed simultaneously at four frequency bands over very low input power ranges. Figure 12.b presents different multi-tone excitations scenarios. Considering two cases; single-tone excitation at each frequency band (4 tones totally) and 12-tone excitation at each frequency band (48 tones totally), black line, and blue line, respectively. In both scenarios, the input power level of each tone is only 0.5 $\mu W\ (-33\ dBm)$ while the output DC power of 9 $\mu W$ and 0.3 $\mu W$ are generated with 48 tones and 4 tones, respectively. Therefore, the output DC power with 12 simultaneous tones in each frequency band (48 tones totally) is 30 times more than 4-tone excitation. It is evident that the QBEH sensitivity is significantly enhanced at very low input power levels. This is due to the rectification capability over a broad input power range and at multiple frequency bands, resulting in effective RF combining.

Fig. 12.c exhibits that the rectifier turns on at a low input power of nearly $-50\ dBm$ with 48-tone excitation, whereas with 4-tone excitation the required turn-on input power is $-40\ dBm$. Hence, using concurrent multi-tone excitation in each frequency band of the QBEH, considerably improves the sensitivity. As can be seen in Fig. 12 (a), the proposed EH can deliver 150 $\mu W$ power with only 5 $\mu W$ input power per each tone with 12-tone excitation at each frequency band (totally 48 tones). This output power is sufficient to supply sensors with higher power consumption [45]

### E. Practical Case Study

Fig. 13 illustrates the output DC power of the QBEH circuit as a function of input power in comparison with other published papers under the same condition. In this figure, a criterion is defined to fairly compare the sensitivity of the proposed QBEH with other works. This criterion is sensor turn-on threshold power (STTP) which is marked in Fig. 13. STTP is the minimum required DC power to turn on a typical sensor which is considered 20 $\mu W$ according to Table III.

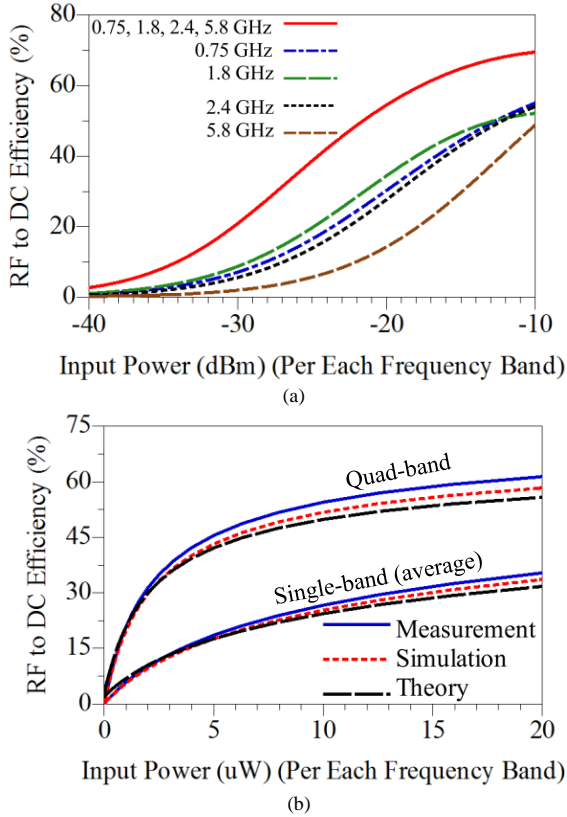

Fig. 11. a) Measured overall efficiency as a function of input power for single-tone and four-tone excitations at input power ranging from $-30\ to\ -10\ dBm$ (1 µW to 100 µW), b) Theoretical, simulation, and measurement results of the overall efficiency for the proposed QBEH in low power region of 0 to 20 µW ($-17\ dBm$).

The minimum required input RF power to generate $20\ \mu W$ output DC power in other published works are marked in Fig.13. For example, in order to produce $20\ \mu W$ output DC power, rectifiers in [27], [50] and [25] require around 70 µW, 100 µW, and 150 µW input RF power, respectively, whereas the proposed QBEH requires only 40 µW input power. Consequently, the proposed QBEH exhibits high-sensitivity power scavenging capability, which is suitable to turn on sensors with low-level ambient signals. Further, recently reported EH circuits are thoroughly compared in Table IV. According to Fig.13 and Table IV, the proposed QBEH exhibits higher sensitivity and efficiency in comparison with other works. This proves the usefulness of the proposed energy harvesting technique for self-sustainable sensors ($S^3$).

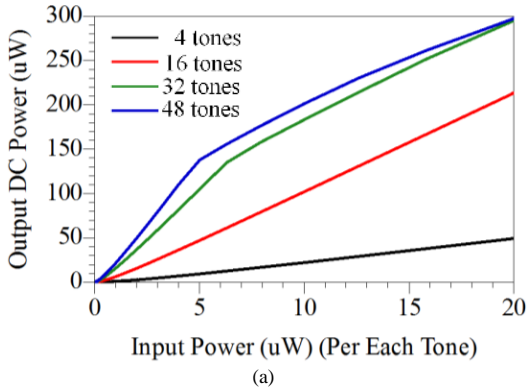

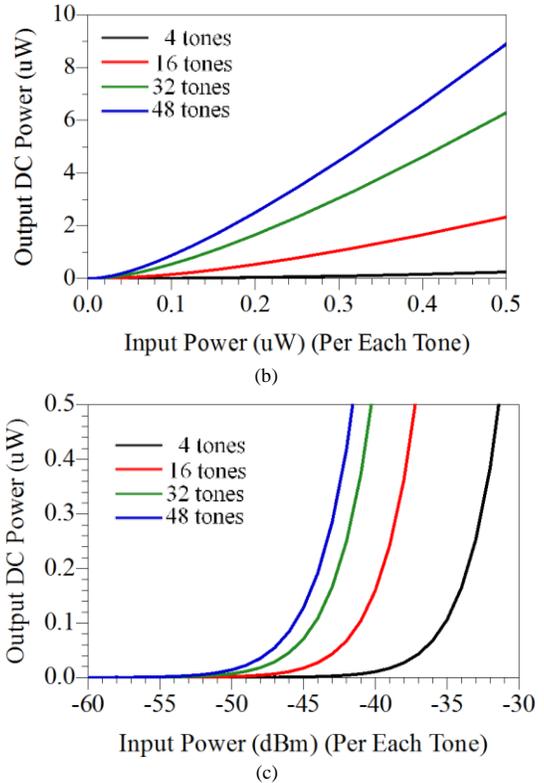

Fig. 12. Simulation results of the output DC power as a function of input power for the proposed QBEH with multi-tone excitations (4 tones, 16 tones, 32 tones, and 48 tones) with the input power of a) 0 to $20 \mu W (-17\ dBm)$, b) 0 to $0.5\ \mu W\ (-33 dBm)$, c) $-60$ to $-30$ dBm, per each tone.

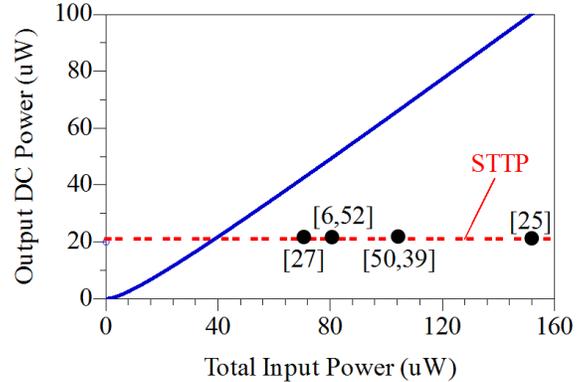

Fig. 13. Measurement results of the output DC power as a function of input RF power for the proposed QBEH in comparison with other works (sensor turn-on threshold power (STTP) is indicated with a dashed line)

## V. CONCLUSION

In this paper, a new technique is presented to realize a compact quad-band energy harvester (QBEH) circuit based on the E-CRLH TL. Simulated and measured QBEH at center frequencies of $f_1 = 0.75\ GHz$, $f_2 = 1.8\ GHz$, $f_3 = 2.4\ GHz$ and $f_4 = 5.8\ GHz$ exhibit a matching (15 dB) bandwidth of 90 MHz, 120 MHz, 60 MHz, and 450 MHz, respectively. The proposed QBEH exhibits a broad matching over a low input power range of $-50$ to $-10$ dBm (0.01 to 100 µW) which makes it a suitable candidate for electromagnetic energy harvesting in urban environments. The proposed QBEH achieved over 55% efficiency at $P_{in} = -20\ dBm\ (10\ \mu W)$ and 70% at $P_{in} = -10\ dBm\ (100\ \mu W)$. Due to the rectification capability over a broad low input power range and at multiple frequency bands, and effective RF combining, sensitivity is improved compared to previously published papers. Precise nonlinear analyses based on the Volterra series and Laplace transformation is presented to obtain minimum input impedance variation of the rectifier over a wide input power range. Theoretical, simulation, and measurement results are in excellent agreement which validate the practicality of the proposed QBEH. Due to multi-band structure, compact size, high sensitivity, and high efficiency, the proposed new quad-band energy harvesting technique has the potential to practically realize a viable energy harvesting solution for self-sustainable sensors ($S^3$), anytime, anywhere. This is of paramount importance for autonomous systems in remote or harsh areas where accessibility is a problem (e.g. mines and large-scale farms). It is the subject of our future work to progress to the prototype stage and integrate QBEH with a multi-band or wide-band antenna and supercapacitor to realize a rectenna. The system will be integrated with sensors and IoT devices to conduct real-environmental measurements.

TABLE IV
PERFORMANCE COMPARISON OF RECENTLY REPORTED EH SYSTEMS.

| Ref. | | This work | [27] 2013 | [25] 2015 | [50] 2018 | [39] 2015 | [51] 2013 | [52] 2017 | [53] 2020 | [6] 2020 |
|---|---|---|---|---|---|---|---|---|---|---|
| Number of bands | | 4 | 2 | 4 | 4 | 2 | 4 | 2 | 3 | 1 |
| Frequency (GHz) | | 0.75<br>1.8<br>2.4<br>5.8 | 1.8<br>2.1 | 0.9<br>1.8<br>2.1<br>2.4 | 0.9<br>1.8<br>2.1<br>2.4 | 0.5<br>0.8 | 0.5<br>0.9<br>1.8<br>2.1 | 2.45<br>5.8 | 1.84<br>2.14<br>2.45 | 5.8 |
| Fractional bandwidth (%) | | 12<br>7<br>3<br>8 | 7<br>6 | 33<br>17<br>10<br>8 | 1<br>1<br>0.5<br>0.5 | 2<br>3 | 10<br>6<br>6<br>14 | 4<br>2 | 2<br>3 | 5 |
| Overall efficiency $\eta_{oc}$ (%) | at $P_{in} = -10dBm$ | 70 | 50 | 60 | 30 | 28 | NA | 35* | NA | 50 |
| | at $P_{in} = -20dBm$ | 55 | 40 | 38 | 20 | 3 | 20 | 15* | 35 | 35 |
| | at $P_{in} = -30dBm$ | 20 | 15 | NA | NA | 0 | NA | 3* | NA | NA |
| Minimum required input power of EH to turn on a sensor ($\mu W$) (Sensor turn-on threshold power (STTP) = $20\mu W$) | | ≅40 | ≅70 | ≅150 | ≅100 | ≅100 | NA | ≅80 | NA | ≅80 |
| Rectifier dimension ($\lambda_g^2$) | | ≅ 1 × 0.2 | 1.8×0.8 | ≅ 0.8 × 0.6 | ≅ 0.2 × 0.2 | ≅ 0.4 × 0.2 | NA | 1.1 × 0.9 | NA | 0.8 × 0.5 |

*Multi-band rectifier without simultaneous multi-tone excitation in each frequency band.

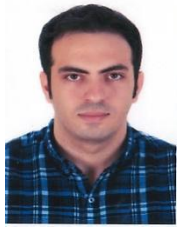

**Rasool Keshavarz** was born in Shiraz, Iran in 1986. He received the PhD degree in telecommunications engineering from the Amirkabir University of Technology, Tehran, Iran in 2017 and is currently working toward an Honorary Appointment position as Visiting Fellow in RFCT Lab at the University of Technology, Sydney, Australia. His main research interests are RF and microwave circuit and system design, sensors, antenna design, wireless power transfer (WPT) and RF energy harvesting (EH).

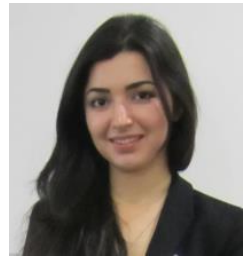

**Negin Shariati** is a Senior Lecturer in the School of Electrical and Data Engineering, Faculty of Engineering and IT, University of Technology Sydney (UTS), Australia. She established the state of the art RF and Communication Technologies (RFCT) research laboratory at UTS in 2018, where she is currently the Co-Director and leads research and development in RF-Electronics, Sustainable Sensing, Low-power Internet of Things, and Energy Harvesting. She leads the Sensing Innovations Constellation at Food Agility CRC (Corporative Research Centre), enabling new innovations in agriculture technologies by focusing on three key interrelated streams; Energy, Sensing and Connectivity.

Since 2018, she has held a joint appointment as a Senior Lecturer at Hokkaido University, externally engaging with research and teaching activities in Japan.

She attracted over $650K worth of research funding over the past 3 years and across a number of CRC and industry projects, where she has taken the lead CI role and also contributed as a member of the CI team.

Negin Shariati completed her PhD in Electrical-Electronic and Communication Technologies at Royal Melbourne Institute of Technology (RMIT), Australia, in 2016. She worked in industry as an Electrical-Electronic Engineer from 2009-2012. Her research interests are in Microwave Circuits and Systems, RF Energy Harvesting, low-power IoT, Simultaneous Wireless Information and Power Transfer, AgTech, and Renewable Energy Systems.

## VI. APPENDIX

To determine the closed-form equation for output DC current ($I_{DC}$), first, Laplace transform is applied to the three linear equations (13) as:

$$\begin{cases} (C_1 R_s s + 1)I_{D1} - I_{D2} - C_1 s V_{j1} - C_1 s V_i = 0 \\ I_{D2} - (C_2 R_L s + 1)I_L = 0 \\ R_s I_{D1} + R_s I_{D2} + V_{j1} + V_{j2} + R_L I_L = 0 \end{cases} \quad (A.1)$$

Laplace transformations of nonlinear equations are very complicated, hence for simplicity, the Volterra series is used in calculations [54].

The nonlinear equations are initially transformed by introducing the concept of the nonlinear transfer function from time domain to multidimensional frequency domain. Then, applying associating variables technique, yields transformation from multidimensions to single dimensions, where all the initial conditions are zero [55], [56].

In general, the Laplace transform of the *n*-order system is equal to [54]:

$$Y_n(s_1, s_2, \dots, s_n) = H_n(s_1, s_2, \dots, s_n) \prod_{i=1}^{n} X(s_i) \quad (A.2)$$

where $Y_n$ is the output Laplace transform, $H_n$ is the Laplace transform of the impulse response and $X$ is the Laplace transform of the input signal. Now, if $H_n$ is known, the output spectrum $Y_n$ can be expressed in terms of the input spectrum $X(s_1), X(s_2), \dots, X(s_n)$.

By inverting $Y_n$ and letting $t_1 = t_2 = \cdots = t_n$, $y_n(t)$ can be obtained for a given *x(t)*.

Please note, in the mathematical literature the quantity $H_n$ is interchangeably termed as Volterra kernel, multidimensional transform or nonlinear transfer function [54].

As an analogy to the linear system, the nonlinear transfer function is defined to be [54]:

$$H_n(s_1,s_2,...,s_n) = \int_{-\infty}^{\infty}...\int_{-\infty}^{\infty} h_n(\tau_1,\tau_2,...,\tau_n) exp\left[-(s_1\tau_1+...+s_n\tau_n)\right] d\tau_1...d\tau_n \quad (A.3)$$

One method of determining the Volterra kernel is harmonic input method. This method is based on the fact that applying harmonic input to the system produces harmonic output.

The harmonic input method states that the *nth*-order nonlinear transfer function can be obtained analytically as the coefficient of $n!.exp(s_1+s_2+\cdots+s_n)t$ in the system output, when the input is [54]:

$$x(t) = exp(s_1 t) + exp(s_2 t) + \cdots + exp(s_n t) \quad (A.4)$$

Now, according to this theory, the Laplace transform of the nonlinear equations (11) and (12) can be calculated. In the following calculations, $x(t) = i_{D1}(t)$ or $i_{D2}(t)$ and $y(t) = v_{j1}(t)$ or $v_{j2}(t)$.

In solving (A. 2), the input $x(t)$ is not bounded and even though the nonlinearity is only second order.

In equation (11), to find the first-order transfer function $H_1(s_1)$, we let $i_{D1}(t) = exp(s_1 t)$ and $v_{j1}(t) = H_1(s_1) exp(s_1 t)$. Substituting the above values of $i_{D1}(t)$ and $v_{j1}(t)$ in (11) and equating the coefficient of $exp(s_1 t)$ on both sides obtains:

$$H_1(s_1) = \frac{1}{I_s\alpha + C_0 s_1} \quad (A.5)$$

To determine the second-order transfer function $H_1(s_1, s_2)$, let:

$$i_{D1}(t) = exp(s_1 t) + exp(s_2 t) \quad (A.6)$$

And

$$v_{j1}(t) = H_1(s_1).exp(s_1 t) + H_1(s_2).exp(s_2 t) + 2H_2.exp(s_1 + s_2) \quad (A.7)$$

For simplicity, the first three terms of Taylor expansion in (11) and (12) are considered. Substituting $i_{D1}(t)$ and $v_{j1}(t)$ values from equations (A. 6) and (A. 7) into (11) and equating the coefficient of $2!.exp(s_1 t + s_2 t)t$, obtains:

$$H_2(s_1,s_2) = \frac{-H_1(s_1).H_1(s_2).\left(\frac{C_0^2}{2V_0}(s_1+s_2)+\alpha^2 I_s\right)}{I_s\alpha + C_0(s_1+s_2)}$$

$$= \frac{-\left(\frac{1}{I_s\alpha+C_0 s_1}\right).\left(\frac{1}{I_s\alpha+C_0 s_2}\right)\left(\frac{C_0^2}{2V_0}(s_1+s_2)+\alpha^2 I_s\right)}{I_s\alpha + C_0(s_1+s_2)} \quad (A.8)$$

However, to find the first two nonzero terms from (A. 7), we achieve:

$$V_{j1}(s_1,s_2) = H_1(s_1).I_{D1}(s_1) + H_2(s_1,s_2).I_{D1}(s_1).I_{D1}(s_2) \quad (A.9)$$

So

$$V_{j1}(s_1,s_2) = G_{11}(s_1) + G_{12}(s_1+s_2).G_{13}(s_1,s_2) \quad (A.10)$$

Where

$$G_{11}(s_1) = \left(\frac{1}{I_s\alpha + C_0 s_1}\right).I_{D1}(s_1) \quad (A.11)$$

$$G_{12}(s_1+s_2) = \frac{\left(\frac{C_0^2}{2V_0}(s_1+s_2)+\alpha^2 I_s\right)}{I_s\alpha + C_0(s_1+s_2)}.\left(\frac{1}{I_s\alpha+C_0 s_1}\right).\left(\frac{1}{I_s\alpha+C_0 s_2}\right) \quad (A.12)$$

$$G_{13}(s_1,s_2) = I_{D1}(s_1).I_{D1}(s_2) \quad (A.13)$$

Now, using the association of variables technique [55], the above quantity translates into one frequency domain. Hence, $G_{11}(s_1)$ and $G_{12}(s_1+s_2)$ are transformed into *s* domain as:

$$G_{11}(s) = \left(\frac{1}{I_s\alpha + C_0 s}\right).I_{D1}(s) \quad (A.14)$$

$$G_{12}(s) = \frac{\frac{C_0^2}{2V_0}s + \alpha^2 I_s}{(2I_s\alpha + C_0 s).(I_s\alpha + C_0 s)} \quad (A.15)$$

In order to translate $G_{13}(s_1,s_2)$ quantity into one frequency domain, $i_{D1}$ is written in the form of:

$$G_{13}(s) = \frac{Ks}{2}.\left[\sum_{i=0}^{N}\sum_{j=0}^{N}\left(\frac{1}{s^2+(\omega_{Li}-\omega_{Lj})^2}+\frac{1}{s^2+(\omega_{Li}+\omega_{Lj})^2}\right)\right] \quad (A.16)$$

Therefore, the Laplace transformation of equations (11) and (12) are:

$$V_{j1} = \left(\frac{1}{I_s\alpha+C_0 s}\right).I_{D1} + \frac{\frac{C_0^2}{2V_0}s+\alpha^2 I_s}{(2I_s\alpha+C_0 s).(I_s\alpha+C_0 s)}.G_{13}(s) \quad (A.17)$$

$$V_{j2} = \left(\frac{1}{I_s\alpha+C_0 s}\right).I_{D2} + \frac{\frac{C_0^2}{2V_0}s+\alpha^2 I_s}{(2I_s\alpha+C_0 s).(I_s\alpha+C_0 s)}.G_{23}(s) \quad (A.18)$$

Therefore, five equations are derived (eqn. (A. a), eqn. (A. 17) and eqn. (33)), with five unknowns which lead to:

$$I_L(s) = \frac{F_2(s).V_i(s) + F_3(s).B(s) + A(s)}{F_1(s)} \quad (A.19)$$

Where

$$F_1(s) = \left(\frac{C_2 R_L s + 1}{C_1 s}\right) - \left(\frac{C_1 R_s s + 1}{C_1 s} + \frac{1}{I_s\alpha + C_0 s}\right).$$
$$\left((-C_1 s)\left(R_s(C_2 R_L s + 1)\left(\frac{1}{I_s\alpha + C_0 s} + 1\right) + R_L\right) - (C_2 R_L s + 1)\right) \quad (A.20)$$

$$F_2(s) = \left(\frac{C_1 s}{I_s\alpha + C_0 s}\right) + C_1 R_s s + 2 \quad (A.21)$$

$$F_3(s) = \left(\frac{-C_1 s R_s(C_2 R_L s + 1)}{I_s\alpha + C_0 s}\right) + R_s(C_1 R_s s + 1)(C_2 R_L s + 1) \quad (A.22)$$

$$B(s) = \frac{\frac{C_0^2}{2V_0}s + \alpha^2 I_s}{(2I_s\alpha + C_0 s).(I_s\alpha + C_0 s)} \cdot \frac{s}{2} \cdot \\ \left(\sum_{i=0}^{N}\sum_{j=0}^{N}\left(\frac{a_i a_j}{s^2 + (\omega_{Li} - \omega_{Lj})^2} + \frac{a_i a_j}{s^2 + (\omega_{Li} + \omega_{Lj})^2}\right)\right)$$ (A. 23)

$$A(s) = \frac{\frac{C_0^2}{2V_0}s + \alpha^2 I_s}{(2I_s\alpha + C_0 s).(I_s\alpha + C_0 s)}$$ (A. 24)

$i_L(t)$ is a time-domain solution of the equation (A. 19) which is driven by the inverse Laplace transform in the form of:

$$i_L(t) = \sum_{i=0}^{N} a_i \cos(\omega_{Li} t) \quad , \omega_{L0} = 0$$ (A. 25)

However, the inverse Laplace transformation of equation (A. 19) is very complicated. On the other hand, the DC term of the load current $a_0 = I_{DC}$, is an essential factor in energy harvesting systems. According to the Residue Theorem, if $I_L(s)$ has a pole of order $m$ at $s=a$, then the residue of $I_L(s)$ at $s=a$ is [57]:

$$r = \frac{1}{(m-1)!} \lim_{s \to a} \left[\frac{d^{m-1}}{ds^{m-1}}((s-a)^m I_L(s))\right]$$ (A. 26)

DC part of the signal $I_L(s)$ is similar to its residue at $s=0$ pole. So, in (A. 26) by setting $m = 1$, $a = 0$:

$$r = I_{DC} = \lim_{s \to 0}[s I_L(s)]$$ (A. 27)

According to (A. 19) and (A. 27) the output current of the load resistance can be calculated.

## VII. Abbreviation

A glossary of used terms is given below:

| Acronym | Description |
| --- | --- |
| $\eta_{effective}$ | Effective efficiency |
| $\eta_{oc}$ | Overall (end-to-end) efficiency of the circuit |
| $\eta_{os}$ | Overall (end-to-end) efficiency of the EH system or rectenna efficiency |
| ADS | Advanced Design System |
| CRLH | Composite right and left-handed |
| CW | Continuous-wave |
| D-CRLH | Dual composite right and left-handed |
| E-CRLH | Extended composite right and left-handed |
| EH | Energy harvester |
| IoT | Internet of Things |
| MTM | Metamaterial |
| QBEH | Quad-band energy harvester |
| QBMN | Quad-band matching network |
| Rectenna | Rectifying antenna |
| $S^3$ | self-sustainable sensor |
| STTP | sensor turn-on threshold power |
| WSN | Wireless sensor networks |